\documentclass{caosp306}
\usepackage{graphicx}
\usepackage{natbib}
\articleNo{1}
\pubyear{2019}
\volume{1}
\volnumber{1}
\firstpage{1}
\received{July 31, 2019}
\accepted{}

\begin{document}
\htitle{Current and future development of Cloudy}
\hauthor{P.A.M. van Hoof {\it et al.}}
\title{Current and future development of the photoionization code Cloudy}
\author{
P.A.M. van Hoof \inst{1}
\and
G.C. Van de Steene \inst{1}
\and
F. Guzm\'an \inst{2}
\and
M. Dehghanian \inst{2}
\and
M. Chatzikos \inst{2}
\and
G.J. Ferland \inst{2}
}
\institute{
Royal Observatory of Belgium, Ringlaan 3, B-1180 Brussels, Belgium
\and
Department of Physics and Astronomy, The University of Kentucky, Lexington, KY 40506, USA
}
\date{July 1, 2019}
\maketitle

\begin{abstract}
The interstellar medium (ISM) plays a crucial role in the cycle of matter in
every galaxy. The gas and dust that is present in the ISM is usually very far
removed from (local) thermodynamic equilibrium, and in some cases may also not
be in a steady-state equilibrium with its surroundings. The physics of this
material is complex and you need a sophisticated numerical code to study it.
For this purpose the open-source photoionization code Cloudy was created.
It models the physical state of the gas and predicts the spectrum that it
emits.

Cloudy is continually being developed to improve the treatment of the
microphysical processes and the database of fundamental data that it uses. In
this paper we will discuss how we are developing the code to improve our
high-density predictions by implementing better collisional-radiative models
for all ions. We will also briefly discuss the experimental mode in Cloudy to
model gas that is not in steady-state equilibrium and present a preliminary
model of recombining gas in a planetary nebula that is on the cooling track.
We finish with a short discussion of how we are speeding up the code by using
parallelization.

\end{abstract}

\section{Introduction}

The space between stars is filled with a very tenuous gas called the
interstellar medium (ISM) which plays a crucial role in the evolution of every
galaxy. This medium is usually irradiated by strongly diluted radiation fields
and is therefore far removed from local thermodynamic equilibrium (LTE). In
some cases it may also not be in a steady-state equilibrium with its
surroundings. The material in the ISM is generally primarily heated by stellar
light, but other energy sources could be shocks, magnetic reconnection, cosmic
rays, radioactive decay, etc. The gas may be ionized, neutral, or molecular
and usually also contains dust grains. The geometry of the gas is often
intricate, making solving the radiative transfer equations difficult. As a
result the physics of this material is complex and you need a sophisticated
numerical code to model the spectrum emitted by this gas.

For this purpose the open-source photoionization code Cloudy was created on 28
August 1978 in Cambridge, UK. The emphasis is on detailed treatment of
microphysical processes, but it also needs simplifying assumptions, such as 1D
spherical geometry\footnote{Christophe Morisset implemented pseudo-3D modeling
  based on Cloudy models using the pyCloudy package \citep{Morisset2013}.}
with simplified radiative transfer, no treatment of shocks, only a very basic
treatment of magnetic fields, and no inclusion of radioactive decay. It models
the physical state of the gas and predicts the spectrum emitted by that gas.
It is the only code that can make a self-consistent model of a photoionized
region and the neutral and molecular regions beyond the ionization front. Such
a code needs a vast amount of atomic and molecular data. Cloudy is continually
being developed to improve the treatment of the microphysical processes and
the database of fundamental data that it uses, with the aim of making the code
suitable for the widest possible range of physical conditions. In this paper
we will briefly discuss our ongoing efforts to improve the code. We will also
discuss the experimental mode in Cloudy to model gas that is not in
steady-state equilibrium and show results from a preliminary model of a
recombining planetary nebula (PN). In the final section we briefly discuss how
we are speeding up the code using parallelization.

\section{The ionization equilibrium}

\begin{figure}
\centerline{\includegraphics[width=0.75\textwidth]{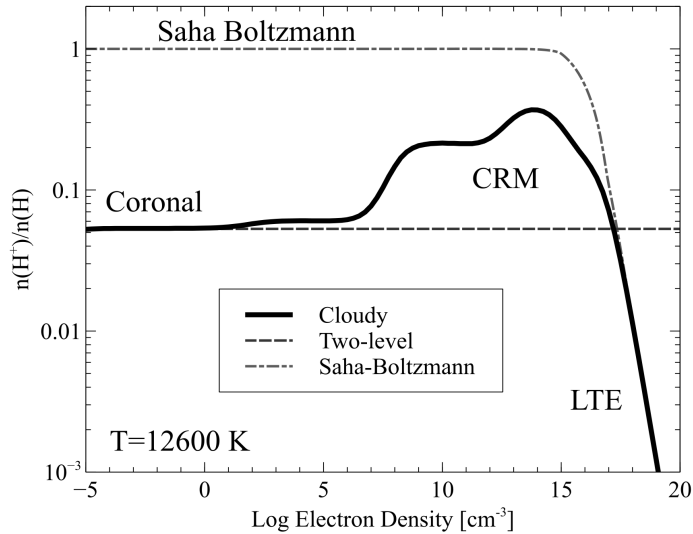}}
\caption{Ionization of hydrogen as a function of density. The solid line is
  the full numerical CRM solution, the dashed line is the ionization predicted
  by the two-level approximation, and the dashed-dotted line is the prediction
  of the Saha-Boltzmann equation. Figure taken from
  \citet{Ferland2017}.}
\label{crmplot}
\end{figure}

To meet our goal of modeling a wide range of physical conditions, Cloudy needs
to be able to calculate the ionization balance over a wide range of densities.
In very low-density gas, the ionization balance can be derived using the
two-level or coronal approximation. Here it is assumed that all the material
is in the ground state, and hence you only need to consider two states: the
ground state of the current and the next ion. This approximation is valid
because the time between two interactions is very long, giving the ion enough
time to relax into the ground state after an excitation or recombination
event. At very high densities, the collisional rates will be very fast,
driving the gas into LTE. This implies that the ionization balance can be
derived from the Saha-Boltzmann equation. Both limits are very easy to model,
however at intermediate densities neither limit is valid and a complex
numerical collisional-radiative model (CRM) is needed to obtain accurate
predictions for the ionization balance. This is shown in Fig.\,\ref{crmplot}.

A CRM solution requires many excited states to be included in the model atom
as the average time between two interactions is too short for the electron to
reach the ground state and excitations or ionizations from excited states
become important. In recent years a considerable amount of effort was
dedicated to prepare Cloudy for this task. Here we will describe the most
important aspects of this work.

\subsection{H- and He-like ions}

For the H- and He-like iso-electronic sequences we use a custom approach for
setting up the model atoms. This is done in such a way that the user can
request an arbitrary number of levels. These can either be {\it l}-resolved or
collapsed into states with the same principal quantum number {\it n}. Such an
approach is possible by exploiting regularities in the atomic data resulting
from the simple level structure of these ions. The original implementation of
the He-like sequence is discussed in \citet{Porter2005, Bauman2005} and the
H-like sequence in \citet{Luridiana2009}. For the 2017 release of Cloudy we
started a comprehensive review of these iso-electronic sequences, mainly
focusing on the accuracy of the collisional data we use \citep{Guzman2016,
  Guzman2017, Guzman2019}. This effort is still ongoing and will result in
further updates in a future release.

\begin{figure}
\centerline{\includegraphics[width=0.70\textwidth]{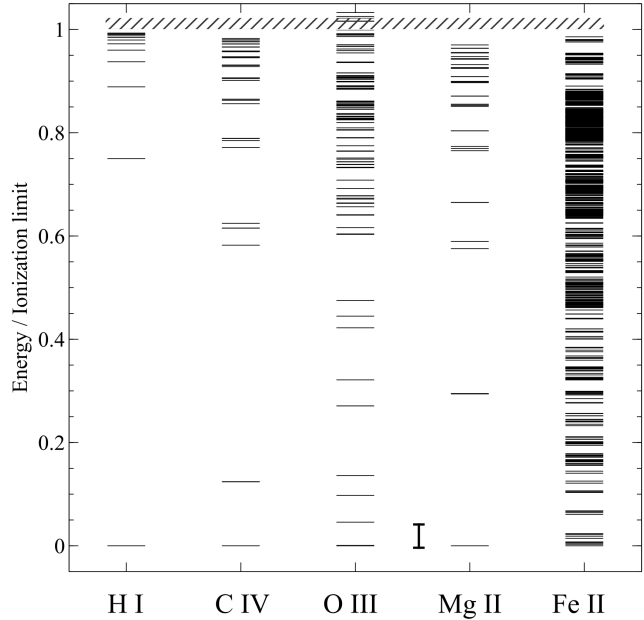}}
\caption{Experimental energy levels taken from NIST ASD \citep{NIST_ASD} for
  some species present in an ionized gas. The energies are given relative to
  the first ionization potential. Figure taken from \citet{Ferland2017}.}
\label{levels}
\end{figure}

The H- and He-like ions are unusual because the first excited level is much
closer to the ionization threshold than to the ground state (see
Fig.\,\ref{levels}). This implies that in photoionized plasmas collisional
excitation is generally unimportant, and the emission line spectra are
produced by recombination. Recombination typically happens in highly excited
Rydberg states, which are tightly coupled to the ionized continuum by
collisional ionization and photoionization and their inverse processes. To
make accurate predictions of the recombination spectrum, a very large model
atom is needed. Cloudy is capable of setting up such a model atom, but the CPU
and memory requirements are currently prohibitive. The reason for that is
twofold. On the one hand, calculating the atomic data is too slow, while on
the other hand solving the equations for the level populations takes too much
time as every level is modeled explicitly. We will tackle the former problem
by caching atomic data or moving them into data files (possibly in Stout
format), rather than calculating them on the fly, and also by using better
optimized algorithms. To alleviate the latter problem, we will implement the
matrix condensation technique pioneered by \citet{Burgess1969} and
\citet{Brocklehurst1970}. This technique is akin to the superlevel technique
often used in stellar atmospheres, which allows a significant reduction in the
number of levels that are modeled, thus significantly speeding up the LU
decomposition needed to solve the level populations. This will allow us to
routinely model large model atoms and improve the accuracy of our
recombination line predictions. This is e.g. needed for the measurement of the
primordial helium abundance. These improvements will be offered in a future
release of Cloudy.

\subsection{More complex ions}

Unlike H- and He-like ions, more complex ions have low-lying excited states.
These can be collisionally excited, even in a photoionized plasma and thus can
be an important source of cooling for the gas. Moreover, these ions can lack
the regularities that H- and He-like ions have, e.g., due to configuration
interaction shifting levels out of position. Because of this, a very different
approach is needed compared to what we described above. We will need to
tabulate the atomic data that are needed to model the atom. The level energies
will normally be taken from experiments, while the transition probabilities and
collision strengths typically come from large-scale computer models.

With the 2013 release of Cloudy we started moving these atomic data into
external data files (until then, the data for a relatively small number of
important lines were hardwired into the code). Initially we adopted the
Chianti database \citep{Dere1997, Landi2012} but we quickly realized that the
scope of this database is insufficient for us to depend on it exclusively
(e.g., it contains insufficient data for low ionization stages). So we decided
to develop our own Stout database for atomic and molecular data. This database
was introduced in the 2013 release of Cloudy \citep{Ferland2013}, but was only
used for a very limited number of ions. With the 2017 release of Cloudy we
moved many more ions to the Stout database, vastly increasing the number of
levels and lines that can be modeled with Cloudy. A more detailed discussion
of these points can be found in \citet{Ferland2017}.

This is the first step towards full CRM treatment of these ions. However, more
work is needed. Many of the entries in the Stout database are what we call
``baseline'' models. These contain level energies and transitions
probabilities taken from NIST ASD \citep{NIST_ASD} but no collisional data. If
for a given transition the collisional data are missing, Cloudy will use the
g-bar approximation \citep{Burgess1992} which is now widely viewed as
unreliable, but is still better than having no data at all. To improve this
situation, we have started a project to update the baseline models using the
latest level energies and transition probabilities from NIST ASD and combining
those with electron impact collisional data from the OPEN-ADAS
database\footnote{http://open.adas.ac.uk/}. This should vastly improve the
quality of the predictions for these model atoms.

Solving the level populations and the ionization balance are however still not
coupled in the sense that ionization from excited states is not considered for
these complex ions (with only a handful of exceptions that are currently
hardwired into the code). We aim to improve this situation in collaboration
with the University of Cambridge by including both collisional ionization and
photoionization processes from metastable states. All these improvements will
be offered in a future release of Cloudy.

\section{Time-dependent modeling}

For quite some time Cloudy has been able to model time-steady non-equilibrium
dynamical flows \citep{Henney2005, Henney2007}. This capability was later
extended to modeling time-dependent non-equilibrium conditions in the gas
\citep{Chatzikos2015}. This feature is still considered experimental. Here we
apply this method to the problem of a recombining PN.

\begin{figure}
\centerline{\includegraphics[width=0.80\textwidth]{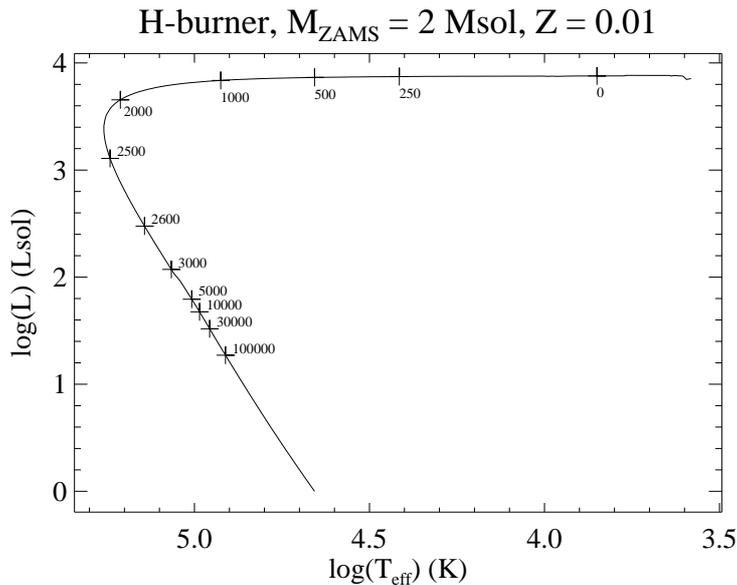}}
\caption{Evolutionary track from \citet{MB2016}. Ages in years since the star
  left the AGB are marked on the track.}
\label{track}
\end{figure}

When an intermediate-mass star nears the end of its life, it will experience
thermal pulses during the asymptotic giant branch (AGB) part of its evolution.
At this time the energy generation is via hydrogen shell burning, punctuated
at regular intervals ($10^4$--$10^5\,$yr) by helium flashes followed by short
phases of quiescent helium burning. During these flashes, the remaining
envelope of the star is removed in a superwind. When the envelope is nearly
depleted, the star starts to heat up and ionizes the material that was ejected
during the AGB phase, forming a new PN. Eventually the hydrogen burning shell
runs out of fuel and the nuclear burning will shut down. At this time the
luminosity of the star will quickly go down while it also will start to cool,
eventually turning the star into a new white dwarf. This part of the evolution
is called the cooling track.

When the star is on the cooling track, the PN will receive increasingly less
ionizing photons. This implies that the outer parts of the PN can no longer be
ionized and will start to recombine (e.g., \citealp{vanHoof2000}). This is a
non-equilibrium process that we can model with Cloudy. Here we present
preliminary results. To have a realistic evolution of the luminosity as a
function of time, we use the 2\,M$_\odot$ hydrogen-burning evolutionary track
from \citet{MB2016} with {\it Z} = 0.01. This track is shown in
Fig.\,\ref{track}.

\begin{figure}
\centerline{\includegraphics[width=0.80\textwidth]{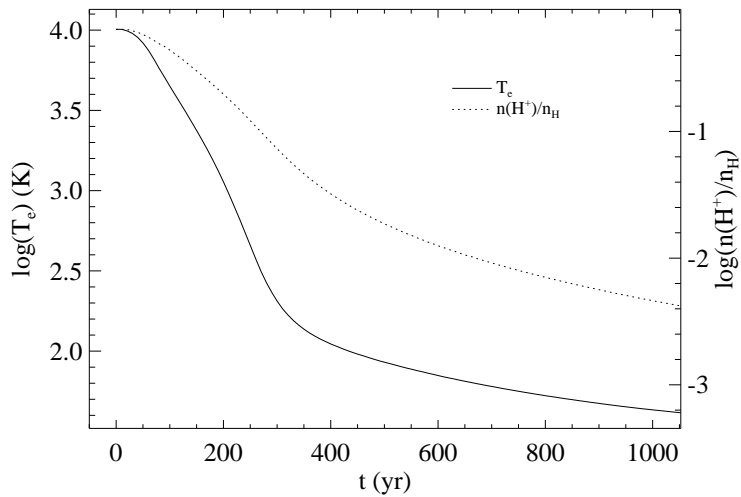}}
\caption{The evolution of the electron temperature and hydrogen ionization
  fraction as a function of time at a radius of $4\times10^{15}$\,m.}
\label{evol1}
\end{figure}

We will start the Cloudy model when the star is at the maximum temperature,
{\it T}$_{\rm eff}$ = 181.5\,kK. This point is reached after 2404\,yr. It is
currently not possible to vary the stellar temperature, so we assume that it
stays constant at this value during the evolution. We model the spectrum of
the star using the H-Ca grid of atmosphere models by \citet{Rauch1997}. The
evolution of the luminosity as a function of time will be given by the
evolutionary track, starting at a luminosity of 2455\,L$_\odot$. The gas is
assumed to have a constant hydrogen density of 1000\,cm$^{-3}$, a reasonable
value for an evolved PN. The inner radius is set at $1.135\times10^{15}$\,m,
which is the distance the gas would have traveled in 2404\,yr assuming a
canonical expansion velocity of 15\,km\,s$^{-1}$. The outer radius is set at
$5.395\times10^{15}$\,m, chosen to coincide with the ionization front at the
start of the simulation. This implies an ionized mass of 0.75\,M$_\odot$. The
gas is assumed to contain graphite dust \citep{Martin1991}.

\begin{figure}
\centerline{\includegraphics[width=0.80\textwidth]{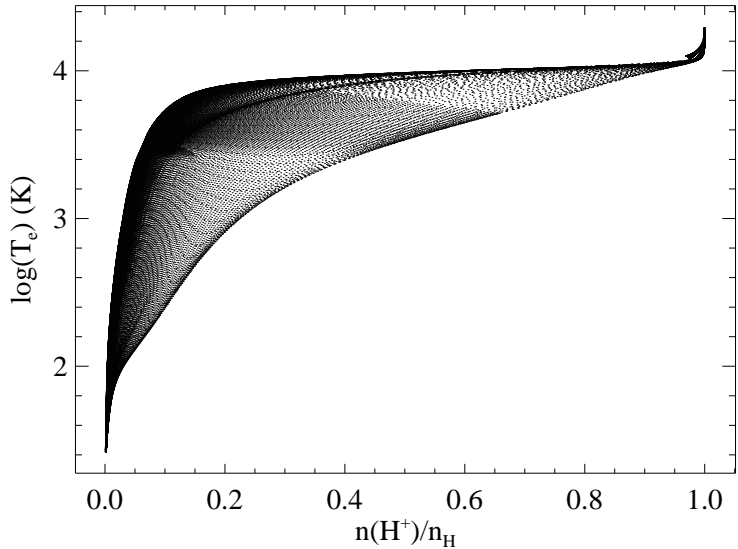}}
\caption{The electron temperature as a function of hydrogen ionization
  fraction is shown for every depth point of every time step of the model.}
\label{evol2}
\end{figure}

The results of the model are shown in Figs.\,\ref{evol1} and \ref{evol2}. The
first shows that the electron temperature drops very quickly: the decrease
from 10\,116\,K to 1000\,K happens in less than 210\,yr, while the hydrogen
ionization fraction lags behind. When the electron temperature reaches
1000\,K, hydrogen is still $\approx 18$\,\% ionized. This can be seen even
more clearly in Fig.\,\ref{evol2} where we show the electron temperature as a
function of the degree of ionization for all depths and every time step. From
this it becomes clear that in a recombining PN it is possible to observe
partially ionized gas at remarkably low temperatures: at 50\,\% ionization,
the electron temperature can be as low as 3410\,K, at 20\,\% ionization, as
low as 815\,K, and at 10\,\% ionization as low as 250\,K!

Roughly similar behavior can be observed for O$^+$, while He$^+$ and N$^+$
also follow this trend, but with less extreme temperature lows. On the other
hand, ions with a higher ionization potential, such as He$^{2+}$, C$^{2+}$,
N$^{2+}$, and O$^{2+}$, do not show this behavior. Their recombination
timescales are much shorter and they will have recombined before the gas has
had the chance to cool by a significant amount. So in a recombining PN, you
can have very cold gas that still has appreciable amounts of H$^+$, He$^+$,
N$^+$, and O$^+$, but not many other ions (C$^+$ is also present, but that has
a lower ionization potential and is still actively photoionized). This gas
will still emit radiation due to the progressing recombination of these ions.
These conclusions qualitatively agree with the earlier studies of
\citet{Ferland1981} who were modeling a nova shell and \citet{Binette1987} who
were modeling an active galaxies.

It is well known that evolved PNe, such as NGC 6720 (the Ring nebula) and NGC
7293 (the Helix nebula), can have a large number of dense, molecular knots
embedded in the ionized gas \citep{Matsuura2009}. The origin of these knots is
still debated. In \citet{ODell2007, vanHoof2010, VandeSteene2015} it has been
proposed that these knots are not remnants of an earlier phase of evolution,
but form in recombining gas when the star is on the cooling track. This leaves
only one to two thousand years to form the knots in the case of NGC 6720
\citep{ODell2007}. They would form as a result of an instability. When the
recombining gas is cold, its internal pressure is low, making it susceptible
to compression, e.g., by recombination radiation. The preliminary models
presented in this paper show that this proposal could be viable. The
recombining gas indeed cools on a very short timescale, but stays partially
ionized, so that it still emits recombination radiation. This agrees with the
assumptions made in \citet{vanHoof2010}.

\section{Parallelization}

Modeling large model atoms as well as doing time-dependent modeling require a
lot of work to be done by the CPU. To ease that burden, it is imperative that
the code makes efficient use of the hardware resources that are available.
This can be improved either through using SIMD (Single Instruction Multiple
Data) instructions (also called code vectorization), or by using multiple
cores in parallel since modern computers have more and more cores available.

In recent years an effort has been made to vectorize Cloudy. One way of
achieving this is to simplify loops such that the compiler can recognize the
potential to use SIMD instructions and generate these automatically while
compiling. For this the loop needs to carry out only one or two simple tasks,
otherwise the compiler will quickly be overwhelmed by the complexity and
compile it using scalar instructions. So to achieve automatic vectorization,
it may be beneficial to split one complex loop into multiple simpler ones.
This task has been carried out in CPU-critical parts of the code.

In addition, we also created code primitives that were explicitly written to
use SIMD instructions. These include routines to carry out reduction loops and
vectorized transcendental functions (which apply transcendental functions to
an array of arguments). Using code vectorization resulted in a moderate
speedup of the code (roughly 20\%). There likely is a potential to find
further optimizations in the future. An added benefit is that any code that
can be vectorized automatically by the compiler can also be parallelized
automatically by the compiler using OpenMP.

Using multiple cores through parallelization promises more substantial time
savings, but is also more complex to implement. In the past we have
implemented parallelization at the highest level by calculating different
models on different cores. This method is highly efficient, but can only be
used in grid or optimizer runs. For a single model different methods are
needed. You can use parallelization on a low level by parallelizing individual
loops as was already discussed above. This approach also opens the possibility
of using graphics processors (though one drawback is that these are largely
optimized for single-precision floating point math, while Cloudy often uses
double-precision variables). Another possibility is mid-level parallelization,
e.g. by solving the level populations of different ions on different cores, or
by solving different grain size bins on different cores. These are ideas that
still need to be tested and implemented in the future.

\section{Concluding remarks}

The interplay of physical processes in photoionized and photo-dissociation
regions is very complex. Modeling these processes requires a large-scale code,
and maintaining this requires a continuous effort. This includes adding the
latest atomic and molecular data sets, implementing new and improved theory
describing the physical processes, implementing faster and better numerical
methods, and of course fixing bugs. In this paper we have highlighted a couple
of the projects that we have recently undertaken, are in the process of
implementing, or are planning to start in the future. These will allow Cloudy
to make more accurate predictions. They will also make the code faster and
more versatile. We hope that these efforts will improve the usefulness of
Cloudy for our users.

\bibliography{peter_van_hoof}

\end{document}